\documentclass[sigconf,screen, nonacm]{acmart}
\usepackage{caption}
\usepackage{multirow}
\usepackage{xcolor}
\usepackage[normalem]{ulem}

\AtBeginDocument{%
  \providecommand\BibTeX{{%
    \normalfont B\kern-0.5em{\scshape i\kern-0.25em b}\kern-0.8em\TeX}}}

\usepackage{threeparttable}
\usepackage{amsmath}
\usepackage{listings}
\DeclareMathOperator*{\argmax}{arg\,max}

\begin{document}

\title{Mind the Gap: On Bridging the Semantic Gap between Machine Learning and Information Security}

\author{Michael R. Smith}
\affiliation{%
	\institution{Sandia National Laboratories}
	\streetaddress{P.O. Box 1212}
	\city{Albuquerque}
	\state{New Mexico}
	\postcode{43017-6221}
}
\email{msmith4@sandia.gov}
\author{Nicholas T. Johnson}
\affiliation{%
	\institution{Sandia National Laboratories}
	\streetaddress{P.O. Box 1212}
	\city{Albuquerque}
	\state{New Mexico}
	\postcode{43017-6221}
}
\email{nicjohn@sandia.gov}
\author{Joe B. Ingram}
\affiliation{%
	\institution{Sandia National Laboratories}
	\streetaddress{P.O. Box 1212}
	\city{Albuquerque}
	\state{New Mexico}
	\postcode{43017-6221}
}
\email{jbingra@sandia.gov}
\author{Armida J. Carbajal}
\affiliation{%
	\institution{Sandia National Laboratories}
	\streetaddress{P.O. Box 1212}
	\city{Albuquerque}
	\state{New Mexico}
	\postcode{43017-6221}
}
\email{ajcarba@sandia.gov}
\author{Ramyaa Ramyaa}
\affiliation{%
	\institution{New Mexico Tech}
	\streetaddress{P.O. Box 1212}
	\city{Socorro}
	\state{New Mexico}
}
\email{ramyaa@cs.nmt.edu}
\author{Evelyn Domschot}
\affiliation{%
	\institution{New Mexico Tech}
	\streetaddress{P.O. Box 1212}
	\city{Socorro}
	\state{New Mexico}
}
\email{eva.domschot@student.nmt.edu}
\author{Christopher C. Lamb}
\affiliation{%
	\institution{Sandia National Laboratories}
	\streetaddress{P.O. Box 1212}
	\city{Albuquerque}
	\state{New Mexico}
	\postcode{43017-6221}
}
\email{cclamb@sandia.gov}
\author{Stephen J. Verzi}
\affiliation{%
  \institution{Sandia National Laboratories}
  \streetaddress{P.O. Box 1212}
  \city{Albuquerque}
  \state{New Mexico}
  \postcode{43017-6221}
}
\email{sjverzi@sandia.gov}
\author{W. Philip Kegelmeyer}
\affiliation{%
	\institution{Sandia National Laboratories}
	\streetaddress{P.O. Box 1212}
	\city{Albuquerque}
	\state{New Mexico}
	\postcode{43017-6221}
}
\email{wpk@sandia.gov}

\renewcommand{\shortauthors}{Smith and Johnson, et al.}

\begin{abstract}
Despite the potential of Machine learning (ML) to learn the behavior of malware, detect novel malware samples, and significantly improve information security (InfoSec) we see few, if any, high-impact ML techniques in deployed systems, notwithstanding multiple reported successes in open literature.
We hypothesize that the failure of ML in making high-impacts in InfoSec are rooted in a disconnect between the two communities as evidenced by a semantic gap---a difference in how executables are described (e.g. the data and features extracted from the data).
Specifically, current datasets and representations used by ML are \emph{not} suitable for learning the behaviors of an executable and differ significantly from those used by the InfoSec community.
In this paper, we survey existing datasets used for classifying malware by ML algorithms and the features that are extracted from the data.
We observe that: 1) the current set of extracted features are primarily syntactic, not behavioral, 2) datasets generally contain extreme exemplars producing a dataset in which it is easy to discriminate classes, and 3) the datasets provide significantly different representations of the data encountered in real-world systems.
For ML to make more of an impact in the InfoSec community requires a change in the data (including the features and labels) that is used to bridge the current semantic gap.
As a first step in enabling more behavioral analyses, we label existing malware datasets with behavioral features using open-source threat reports associated with malware families.
This behavioral labeling alters the analysis from identifying intent (e.g. good vs bad) or malware family membership to an analysis of which behaviors are exhibited by an executable.
We offer the annotations with the hope of inspiring future improvements in the data that will further bridge the semantic gap between the ML and InfoSec communities.

\end{abstract}


\maketitle

\section{The Promise of Machine Learning}
Recently, machine learning (ML) has reached previously unseen levels of success particularly in deep learning (DL), a class of ML algorithms that use multiple layers in a neural network.
Current advances in DL have achieved state-of-the-art results in many application areas including computer vision \cite{Szegedy2017_AAAI, Hu2018_CVPR}, medical diagnosis \cite{Erickson2017_RG}, machine translation \cite{Vaswani2017_NIPS, Radford2019_TechReport}, and playing games \cite{Mnih2015_Nature, MuZero2019}.
As such, ML \emph{promises} the ability to achieve expert-level performance making previously unobtainable results possible.
For malware detection, ML \emph{promises} to go beyond signature-based techniques by learning the behaviors of malware that is capable of detecting previously unseen malware.
Prior success in ML demonstrates the possibilities and creates a certain amount of hype for similar success.
The successful application of ML in malware detection stands poised to reduce manual labor by orders of magnitude, reduce errors, and work at scales and speeds previously unobtainable.

Practical success in ML-based malware detection, however, has not yet been realized despite reported success in ML and information security (InfoSec) open research communities.
The discrepancy between the reported performance by ML papers and practical results has made many inside the InfoSec community wary of using ML.
The output from the ML algorithm is often disregarded or manually verified---countering the benefit of using ML.
There are many perils to using ML in detecting malware, including:
1) overconfidence in the model based on unrealistic reported results when evaluating the ML model,
2) distrust in the black box models leading to unnecessary work of verifying the output, and
3) vulnerability to intentional adversarial attacks on a large input space.
These caveats are not singular to detecting malware, but are particularly acute given the high consequences of misclassifying a malware sample.

We hypothesize that this discrepancy stems from a disconnect between the ML and InfoSec communities and is observable by a semantic gap (the difference between two descriptions of an object \cite{Hein2010_GAP}) in describing executables.
With the end goal of detecting malware based on its behaviors, we survey the existing literature on the current datasets that are used by ML models to classify malware highlighting this semantic gap. 
Bridging this gap is especially important to ML as the culture generally emphasizes demonstrated performance improvements on benchmark datasets (sometimes over an increase in knowledge) \cite{Sculley2018_ICLR}.
The current culture has driven significant improvements by orders of magnitude but places complete dependence on an appropriate benchmark dataset.
Misaligned datasets result in poor performance in practical scenarios.

In highlighting this semantic gap, we do not imply that previous work is irrelevant.
Rather, due to the current representations used by ML, we find that most approaches are an efficient method for learning signatures.
These have the same caveats of being easily by-passed with relatively simple methods while a possessing a \emph{limited} ability to extrapolate to unseen malware samples.
We put forth that current datasets are limited in their ability to perform behavior-based malware detection in two primary ways:
\begin{enumerate}
	\item \emph{Not being representative of data that will be encountered in deployed scenarios}.
	A commonly employed approach selects samples that several signature-based detectors agree upon---biasing the data to the most obvious examples as defined by signature-based detectors.
	It follows that ML algorithms will learn to replicate signature-based detectors.
	Additionally, due to legal constraints, obtaining large amounts of goodware to distribute is challenging.
	\item \emph{Feature representation is not behavioral.}
	Sharing live malware samples is dangerous.
	Even if live malware is provided, ML algorithms are not designed to ingest them and some processing is required.
	Extracting the proper representation over a large corpus of executables is challenging and time-consuming.
	As a result, simpler features are used and relevant behavioral information is lost.
\end{enumerate}
We suggest that to improve the adoption of ML-based malware classification, we first need to bridge this semantic gap by aligning the data used by ML algorithms with the goals of the InfoSec community.
As a first step, we provide a method for annotating behaviors in malware samples using open resources and provide behavioral annotations for the Microsoft Malware Classification Challenge dataset \cite{Ronen2018_MSKaggle}.
This change in labeling alters the focus from discriminating malware to identifying the key behaviors.
In our view, properly crafted and labeled datasets for ML promises to identify behavioral characteristics and move beyond signature-based detections as ML will optimize behavioral identification.

\section{Preliminaries}
In this section, we provide a brief overview of ML, caveats for its success, and a brief overview of program analysis (PA) techniques used by InfoSec teams to determine the intent of an executable.

\subsection{Machine Learning Background}
\label{sec:MLbackground}
In this paper, we focus on \emph{supervised} ML that learns by example from labeled data points.
We denote the inputs or features as $X$ (where $X$ can be represented as a vector and each individual component can be accessed by subscripts $X_i$) and the labels or outputs as $Y$.
Observed variables are represented in lower-case.
Therefore, the $i^{th}$ observation of $X$ is written as $x_i$ which can be a vector or a scalar.
Informally, the goal of supervised ML is, given an input vector $X_i$, predict an output $\hat{Y}_i$ that matches the actual label $Y_i$.

We follow notation from Friedman et al. \cite{StatisticalLearning} to more formally describe supervised ML.
Suppose that our data was generated from a statistical model
\begin{equation}
\label{eq:dataModel}
Y = f(x) + \epsilon
\end{equation}
where the noise $\epsilon$ has $E(\epsilon) = 0$ independent of $X$.
The goal of an ML algorithm, then, is to find an approximation $\hat{f}(x)$ to $f(x)$ that preserves the predictive relationship between $X$ and $Y$.
The approximation $\hat{f}(X)$ is learned from a training set $\mathcal{T}$ of $N$ observed input-output pairs $(x_i, y_i)$, $i = 1,\dots,N$.

An ML algorithm can modify the input-output relationship $\hat{f}(x_i)$ in response to the difference between the prediction $\hat{f}(x_i)$ and the observation $y_i$. 
Each learning algorithm has an associated set of parameters $\theta$ that can be modified to alter $\hat{f}(x)$.
A simple example is a linear model where $f(x) = x^T\beta$ and $\theta = \beta$.
The values of $\theta$ are found by minimizing a loss function $L$ that measures the ``goodness'' of the model fit as a function of $\theta$ ($L(Y,f_\theta(X))$).
The loss function can take a number of different forms and has a significant impact on the results and the optimization of the learning algorithm.
Many ML algorithms are \emph{maximum likelihood estimators} assuming that the most likely values for $\theta$ are those that provide the largest probability of the observed values:
$$
\hat{\theta} = \argmax_{\theta \in \Theta} \sum^N_{i=1}\log Pr(y_i|x_i, \theta).
$$

For example, one loss function minimizes the residual sum-of-squares ($RSS$) or, another, the cross-entropy ($CE$) loss when $Y$ is a vector of $K$ possible classes:
\begin{gather*}
RSS(\theta) = \sum^N_{i=1}(y_i-f_\theta(x_i))^2\\
CE(\theta) = \sum_{i=1}^N \log p_{y_i,\theta}(x_i)
\end{gather*}
where $p_{y_i,\theta}(x_i)=Pr(Y=Y_k|X=x), k = 1,\dots,K$ for the conditional probability of each class $k$ given $X$.

ML algorithms minimize the loss on an observed training set $\mathcal{T}$ (\emph{training error}), however, the goal is to minimize the error on unobserved data points (the \emph{test} or \emph{generalization error}).
The expected generalization error can be decomposed:
\begin{align}
Err(x) & = E[(Y-\hat{f}(x))^2]\nonumber\\ 
& = (E[\hat{f}(x)]-f(x))^2 + E[(\hat{f}(x)-E[\hat{f}(x)])^2] + \sigma^2 \label{eq:err}
\end{align}
which is a sum of the bias, variance, and the irreducible error.
The irreducible error represents the inherent noise in the data ($\epsilon$ in Equation \ref{eq:dataModel})---no matter how good the model is, there will be some amount of error.
The bias is the difference between the average model prediction and the actual value.
High bias refers to models that focus less on the training data and possibly oversimplifies the model.
High variance models, on the other hand, focus more on the training data and possibly result in overly complex models.
As the complexity of a model increases, the training error tends to decrease.
The performance on training data is usually not a good indicator of how the model will \emph{generalize} or how well it will perform on new data points.
Thus, a trade-off between bias and variance is needed to achieve a model that generalizes the best to test data.
There are a number of methods for appropriately trading off between bias and variance.
A commonly used approach to estimate the generalization error is $k$-fold cross-validation.
In $k$-fold cross-validation, $\mathcal{T}$ is divided into $k$ partitions and each is used as a validation set while the model is trained on the remaining $k-1$ partitions.
The average performance over the $k$ partitions is then reported.
For more details, see an ML overview such as Friedman et al. \cite{StatisticalLearning}.

We emphasize that there is an explicit dependence between the quality of $\mathcal{T}$ and the quality of $\hat{f}(x)$.
In practical situations, ML typically requires large amounts of effort in gathering, cleaning, and processing data.
An observed data point $x_i$ needs to be represented in a format that an ML model can operate on---most ML algorithms operate on vector representations.
However, many interesting problems are \emph{not} easily represented as vectors without throwing away significant amounts of information.
If the representation of the data does not contain the information required for the question that is being asked (e.g. is the behavior of this executable benign or malicious?) then this falls within the irreducible error ($\sigma^2$ from Equation \ref{eq:err}) that cannot be overcome.

\subsection{Overlooked Caveats of ML Successes}
\label{sec:MLcaveats}
ML has been successfully applied in several different domains.
However, this success is usually built on decades of previous research and understanding that is lacking in InfoSec.
For example, the success of convolutional neural networks (CNNs) \cite{Lawrence1997_TransNN} in image and speech processing builds on decades of research in signal processing and how to properly represent the data.
The convolution is an mathematical operator that expresses the amount of overlap of one function over another function and can be thought of blending one function with another.
For example, a Gaussian function can be used as a convolution filter to produce a Gaussian blur in image processing.
The convolutions in CNNs, are a codification of convolutions where the learned function is based on the data instead of being explicitly defined.
One key reason for the success of convolutions is their ability to be translationally invariant which is inherently important in object recognition where an object may be anywhere in the image.
An analogous operator in the binary executable analysis domain does not yet exist. 

Another key aspect of the success of ML in other domains is large amounts of labeled, real-world relevant datasets.
Li revolutionized the computer vision field and object detection by providing labels for relevant images \cite{imagenet_cvpr09}.
Corresponding datasets do not yet exist for malware detection.
Further, ML models do not always learn the intended concepts.
ML algorithms tend to learn signatures, even when the semantic data is present.
For example, CNNs are biased towards learning texture rather than shapes and objects as humans perceive \cite{Geirhos2019_ICLR, sinha2020curriculum} making them susceptible to adversarial attacks.

Adding to these issues, the data in InfoSec applications is significantly different from other domains.
The data lacks proximity relationships, continuity, and ordinality that are present in many other domains and assumed by many ML algorithms.
For example, pixel values of 123 and 122 are close in value and neighboring pixels would have an assumed proximal relationship.
In executables, code blocks can jump to various locations in the binary and values next to each other in numerical space can have significantly different meanings.
Further compounding this issue is the fact that in general the number of malware samples is significantly outnumbered by the number goodware samples which has been shown to exacerbate other issues in ML algorithms \cite{Smith2014_MLJ}.
The combination of these issues makes simply applying current ML algorithms efficiently difficult. 

\subsection{Program Analysis}
\label{sec:PA}
processes that are used to reason about the behavior of a computer program.
In its base form, PA is ultimately interested in program optimization and correctness such as compiler optimization.
There are several research threads revolving around PA.
We highlight a subset of areas related to extracting features that could be used as input to ML algorithms.

\subsubsection{Syntax vs Semantics}

In PA, a distinction is made between the \emph{syntax} and the \emph{semantics} of a program \cite{Hennessy1990_Semantics} similar to the syntax and semantics found in natural languages, where the syntax is often described by a grammar (e.g. nouns, verbs, and proper ordering) and the semantics are the meaning behind the sentence(s). 
For programs, syntax is concerned with the form of expressions that are allowed (i.e. the sequences of symbols that are accepted by a compiler/interpreter). 
Semantics describe the effect of executing syntactically correct expressions (behavior).
To determine the semantics of a program, a definition of syntax is required, at least at an abstract level.
Identifying syntax is much easier than semantics. Consequently, syntactic features are commonly used in ML.

In the context of ML-based malware detection, the ability to detect previously unseen malware and possible zero-day attacks implies that the ML model is able to move beyond syntactic discrimination (like signature detection) to capture behaviors in the semantics.
As shown in Section \ref{sec:MLbackground}, the performance of an ML model is dependent on the training data.
If the training data is syntactic then expecting an ML model to learn the semantics is unreasonable.

\subsubsection{Static Analysis Techniques}

In static analysis, a program is analyzed in a non-runtime environment.
The analysis is generally performed on a version of the source code, byte code, or application binaries. 
Static analysis is used frequently for optimization such as dead code
elimination and for verification such as identifying potentially vulnerable code and run-time errors.
Generally, static analysis \emph{approximates} all possible executions of a program through abstract interpretation or data-flow analysis. 

There are several static analyses techniques that capture semantic information.
However, many datasets used for ML favor syntactic features that are easier to capture.
Data flow analysis (DFA) is a frequently used technique that collects information about the possible states (e.g. how variables are used throughout a program) at various points in a program \cite{Sharir1978_DFA}.
DFA constructs a control flow graph (CFG) that represents the program.
Each node in the CFG often represents a basic block or a sequence of consecutive instructions where control can only enter at the beginning of the block and leaves at the end of the block.
Directed edges in the graph represent jumps between one basic block to another.
Kildall's method is a common method for performing DFA where an equation for each node is derived and each equation in the graph is iteratively solved, propagating inputs and outputs until the system converges \cite{Kildall1973_ACM_SIGACT}.
For example, in forward flow analysis, the equations are in the form:
\begin{align}
out_b &= trans_b(in_b)\nonumber\\
in_b &= join_{p\in pred_b}(out_p)\nonumber
\end{align}
where $trans_b$ is the transfer function of block $b$ representing how $b$ affects the state of the program and $join$ combines the exit states of the predecessors of $b$ as the entry state of $b$.
Once the equations are solved, the entry and exit states of the basic blocks can be used to derive certain properties about the program.
The CFG is a more semantic representation of the executable, yet how to represent the CFG suitable for an ML algorithm without loosing semantic information is non-trivial.

Abstract interpretation \cite{Cousot1977_ACM_SIGACT, Cousot2014_ACM_CSL-LICS} is a theoretical framework to formalize the approximation of computing abstract semantics.
Here semantics refer to a mathematical characterization of possible behavior of a program.
The most precise semantics describe accurately the actual execution of a program and are called concrete semantics. 
Small-step, or structural oriented, semantics \cite{Plotkin1981_note} describe a program in terms of the behaviors of its basic operations.
The behavior of a program is a current state (program point and the environment) given a starting state and series of operations.
For example, consider the simple code below.
\begin{lstlisting}
	1: n=0
	2: while n < 500 do
	3:   n = n+1;
	4: end
	5: exit
\end{lstlisting}
Analyzing the program would yield:
\begin{align}
	&<1, n \Rightarrow \Omega> \rightarrow <2, n \Rightarrow 0> \rightarrow <3, n\Rightarrow 0> \rightarrow <4, n \Rightarrow 1>\rightarrow\nonumber\\
	&<2, n \Rightarrow 1> \rightarrow <3, n\Rightarrow 1> \rightarrow <4, n \Rightarrow 2>\nonumber \dots <5, n \Rightarrow 500> \nonumber
\end{align}
Operational semantics, such as small-step semantics, combine logical conclusions about program syntax in order to derive semantic meaning.
Assuming the interpretation of syntax is correct, this also allows for the construction of proofs about program behavior.

Big-step, or natural, semantics \cite{Kahn1987_STACS}, like small-step semantics, define basic components to describe the semantics of a program.
Rather than using the basic operations like small-step, big-step analytics defines the semantics of functions.
Both small and big-step semantics aim at the same purpose, but go about it in different ways.
More pertinent to malware classification, both are techniques that derive semantic meaning from a program and could be looked to as inspiration for features.

Another key static analysis approach over programs is symbolic execution.
Symbolic execution techniques build a mathematical representation of a program based on the input and output of various subroutines or functional blocks \cite{anand2008demand, ma2011directed}.
In this representation, independent variables represent key input values.
Constraint solvers, for example, can then solve for the variables, identifying what kinds of inputs are required for a particular output state \cite{rossi2006handbook, jaffar1987constraint}.
From a vulnerability analysis perspective, this can allow analysts to identify input that can potentially lead to system failure states, which may be exploitable. 
Though powerful, symbolic execution techniques suffer from state explosion proportional to the size and complexity of a given program \cite{kuznetsov2012efficient}.

In addition to the more formal approaches, other static analysis techniques provide disassembly and intermediate-representation of the code (going from binary to machine code).
These provide some possible steps in transitioning from syntactic to more semantic representations.
However, care needs to be taken to ensure that the semantic information is not lost.

\subsubsection{Dynamic Analysis Techniques}

Dynamic analysis analyzes a program by executing it and \emph{precisely} analyzes a single or limited number of executions of a program.
The coverage of dynamic analysis is dependent on the test inputs, which for malware analysis, can be variants of the operating environment.
Often, a subset of the interactions with the underlying operating system are analyzed such as system calls or memory reads/writes.
It is often used to ensure program correctness and finding errors in the code \cite{Meyers2011_ArtSWTesting}. 
We focus on techniques that we see as the most pertinent for ML-based malware detection.

Most dynamic analysis techniques use instrumentation to insert code into a program to collect run-time information.
The instrumentation will vary based on the type information that is desired and the type of code that is available (e.g. source code, static binary, and dynamic binary).
While each has its own merits, we discuss briefly dynamic binary instrumentation as there is no need to recompile or re-link libraries and it can be attached to running processes and minimizes an inadvertent modification to malware behavior.
Most tools track function calls including system calls and capture the input parameters, track application threads, intercept signals, and instrument a process tree.

The output from dynamic analyses has often been heralded by ML practitioners as a means of modeling behavior.
However, because of a lack of context and the challenges outlined
previously, the representations that are suitable for ML often lose the semantic information.

\subsection{Behavioral Malware Classification}
All software has a specific intent or goal.
To achieve these goals, behaviors have to be executed that align with accomplishing them.
\textit{Behavioral-base malware detection} evaluates a piece of
software by its intended actions before it actually executes them, thus providing a natural intersection between the ML and PA communities.

In general, a behavior can be defined as an observable changes to the state of a system such as registry modification, file interaction and handles, and API calls \cite{Mosli2017_AdvDF}.
Previous work has examined a bottom-up approach developing signatures for the behavior of binary code and combining short pieces of code to get a higher level behavior in human-readable format for malware detection \cite{Afzal2020_SCN, Galal2015_CVHT, Liu2011_WorkCDM}.
Mapping to lower-level behaviors, Bayer et al. \cite{Bayer2009_NDSS} create a profile of malware that models objects such as files and registry keys and the operations made on them. 
They then cluster malware according to its behavioral profile

Given that a behavior can be represented in numerous ways a top-down approach may be more appropriate.
One approach examined network traffic and manually assigned high-level behaviors \cite{Deng2018_USENIX} such as `scanning', `scamming', `downloader' etc. based on observed artifacts.
This approach and other similar approaches \cite{Rieck2011_JCS, Yu2018_FITEE} get at assigning behaviors, yet still remain somewhat brittle as signatures are assigned to behaviors.

We propose that mapping malware to their behavior in a comprehensive manner can be facilitated with improved labeling and representation.
Such a dataset will enable the mapping of individual behaviors to sequences of lower-level semantics extracted from binary files.

\section{Motivating Case Studies}
\label{sec:motivating}
To help motivate the disconnect between the ML and InfoSec communities, we walk through a case of using ML to automate identifying malware persistence in registry keys.
We highlight the difficulty in generating an appropriate dataset and extrapolating the results to real-world scenarios.

Briefly, the Registry is a hierarchical key-value database that stores configurations, program settings, and user profiles.
The value is capable of storing commands to execute when the system is loaded and is commonly used for maintaining persistence on the Windows operating system \cite{Microsoft18}.
In addition to system software, malware takes advantage of the the Registry to ensure that it is loaded as needed.
As an example, a key can have the format:
\begin{verbatim}
\HKEY\_LOCAL\_MACHINE\System\...\...\ImagePath
\end{verbatim}
and a value that can take multiple formats such as:
\begin{verbatim}
C:\Windows\System32\svchost.exe -k netsvcs
or
C:/Program Files/Java/jre7/bin/jqs.exe -service -config
C:/Program Files/Java/jre7/lib/deploy/jqs/jqs.conf
\end{verbatim}
The examples represent a path to an executable, but the values are capable of storing many complex data types (e.g., binary data, scripts, etc.).
Thus, even with this relatively simple example, representing this data in a format suitable for ML is non-trivial.

\begin{figure}[t]
\begin{center}
$\bordermatrix{ &Microsoft&EXE&\ldots & Java\cr
                r_1& 0 &  1  & \ldots & 0\cr
                r_2& 1  &  1 & \ldots & 0\cr
                \vdots & \vdots & \vdots & \ddots & \vdots\cr
		r_{n-1}& 0  &   1       &\ldots & 2\cr
                r_{n}& 1  &   1       &\ldots & 0}$
\end{center}
\caption{Simple Example of a Resulting Object-Term Matrix}
\label{fig:terms}
\end{figure}

\subsection{Data Collection \& Parsing}
As with most use cases, collecting data is not challenging, but obtaining labels and properly representing the data is.
Registry data was collected from Windows machines across a corporate network for roughly two years, beginning in October 2012, resulting in approximately 20 million (host, Registry key, timestamp) tuples, with roughly 136,000 unique Registry entries.
Registry data collected from executing publicly-available malware in a sandbox environment produced 200 Registry entries.

The raw Registry data is not suitable for ML algorithms due the variability of the keys.
As there are a finite number of keys, they are represented as a $1$-of-$N$ encoding that indicates which Registry key is being used.
The value portion is more complex and describes what is being executed and from which directory.
Ideally, the value consists of a path and a file that can be parsed into its relative components.
However, in some cases one program will launch another which we want to be able to capture.
This situation occurs in the case of services on the computer, launched using \verb+svchost.exe+, and in a variety of other situations, e.g. launching dynamically-linked library (DLL) files using \verb+rundll32+, or running Java programs using the Java virtual machine. 
For these situations, we developed a parser that finds the launching program (e.g., \verb+svchost+) as well as the program that is being launched.  
Each launching program is parsed according to the expected syntax (e.g., \verb+svchost+ should have a \verb+-k+ flag), and when found, these launching programs constitute another categorical variable.
Additionally, different file types exist which are represented as categorical variables per file type including any associated options (e.g., command-line flags). 

Finally, after performing the aforementioned parsing, the specific folders in a given path are used as terms in a traditional bag-of-words model.
\figurename~\ref{fig:terms} provides an simplified example of the resulting representation.
The resulting data is high-dimensional (over 12,000 terms) and extremely sparse with few unique observations (i.e., the number of unique rows is close to the number of columns).
In order to transform the data into a form more amenable to analysis, Principal Component Analysis (PCA) was performed on the initial representation to reduce the dimensionality of a given set of variables (i.e., features) while preserving as much information about the original space as possible. 

As outlined in the process, several assumptions and trade-offs were made to massage the data into a format suitable for ML including discarding information.
A trade-off had to be made between capturing all possible information and the complexity of training an ML model.

\subsection{Experimental Analysis and Bias}
Given a format suitable for ML, labels are needed to identify which Registry keys are associated with malicious or benign activity.
Initially, we considered any key that occurs on a large number of hosts as benign and those that were modified by the malware as malicious.
Experimentation with this setup resulted in a cross-validated area-under-the-curve (AUC) score of $0.99$.
Performance this high should suggest that the ML problem is too simple and thus will not be practically useful.
Upon closer inspection, the malicious examples came from specific hosts and identifying the malware labels was a simple process.
Further investigation revealed that Registry keys that occur on a large number of systems tend to be associated with programs and drivers in the system space (e.g., in \verb+C:\Windows\system32+).
However, the majority of the malicious keys are associated with the user and program space.
A simple weak indicator that looks for absence of the keywords ``windows'', ``system'', or ``program'' to determine maliciousness provides an AUC of $0.85$. 
Thus, our model inadvertently models system space keys versus other keys
and is not likely to generalize well.

Only labeling keys modified by malware as malicious and all others as benign results in an AUC of $0.96$ for ML and an AUC of $0.53$ for the weak indicator---not significantly better than random.
This result is promising as the gap between ML and a simple indicator increased significantly. However, this correction is likely still optimistic.
Cross-validation tends to be optimistic in general, due to the fact the errors are not independent.
Also, this data is not likely to contain all possible examples of malware that uses legitimate software Registry for persistence.
However, creating a generalizing principle beyond a signature is challenging.
Another confounding factor is that malware can execute behavior that is not malicious.
Malware authors execute benign software to avoid detection, and, thus, make it difficult to derive ground-truth labels.

\subsection{Other Examples}
This paper is not the first to recognize the gap between the research and actual deployments.
Sommer and Paxson \cite{Sommer2010_SSP} point out the discrepancies in network intrusion detection.
They observe that the task of intrusion detection is fundamentally different from other applications, making it more challenging.
They identify six key challenges:
1) ML is better for finding similarities rather than differences,
2) very high cost of classification errors, 
3) a semantic gap between detection results and their operational interpretation,
4) enormous variability in what is ``normal'',
5) difficulties in sound evaluation of the results, and 
6) operating in an adversarial setting.
In the context of detecting malware, other work noted discrepancies particularly with respect to the precision of malware---indicating a large jump in false negatives when deployed in real-world settings stemming from the difference in the proportion of malware and the difficulty on modeling ``normal'' in executables \cite{Smith2018_ICMLA}.

\section{Current Datasets}
\label{sec:datasets}
The quality of a learned ML model is largely dependent on the quality of the training data (Section \ref{sec:MLbackground}) including how representative the dataset is of real-world scenarios, the alignment of the information conveyed in the features with domain-relevant questions, and the amount of mislabeled examples or noisy features.
PA techniques are available that extract several different types of features, yet prior work has mostly used syntactic features which are easy to extract.
In this section, we briefly discuss the importance of benchmark datasets historically in ML research, discuss the challenges in curating a benchmark dataset for malware, and review the current set of datasets.
Despite several attempts, a benchmark dataset for malware classification has yet to be widely adopted and have a high impact for ML-based malware classification.

\subsection{The Utility of Benchmark Datasets}
The progress of any research field depends on reproducible comparisons
between methods to quantify progress on a given task.
For ML, benchmark datasets facilitate comparisons between learning algorithms.
In addition, benchmark datasets drive ML success and guide research in several application areas such as object detection \cite{Deng2009_CVPR}, facial recognition \cite{Phillips1998_IVC}, handwriting recognition \cite{LeCun1998_IEEE}, recommender systems \cite{Harper2015_IIS}, and question and answer systems \cite{Rajpurkar2016_squad}.

As obtaining data for ML problems is difficult for various reasons, benchmark datasets provide a valuable asset that allows research to be conducted that would not be possible without them.
As such, perhaps more than any other domain, a quality dataset is the key for ML success.

A benchmark dataset dictates several important characteristics of the research that use the benchmark dataset.
First, it determines which features are possible based on how the data is represented.
Second, it determines the impact of ML models developed using the data.
If the dataset misrepresents the real-world settings, the ML model will perform poorly despite performing well on the benchmark dataset.
Thus, great care needs to be exercised when developing a benchmark dataset.
Curating a benchmark dataset is easier for certain domains than for others and creating a benchmark dataset for malware is more challenging than others as discussed in the next section.

\subsection{Challenges in Curating a Malware Dataset}

\subsubsection{Dynamic Environment}
Malware classification is a dynamic problem in which the target is constantly changing and evolving.
In ML parlance, this is concept drift where the distribution of the target changes over time from what was used for training \cite{Gama_2014_ConceptDrift}.
In cases with concept drift, performance often degrades and has been shown to be significant in malware detection \cite{Kegelmeyer13}.
In addition to being highly dynamic, malware authors intentionally alter malware to avoid detection using several obfuscation techniques including polymorphic code and garbage code insertion.
In many other domains, the attempt to deceive is not as prevalent.
One danger is that once a dataset is established, malware authors can purposefully alter their malware to subvert an ML model trained on that dataset, rendering it obsolete.

\subsubsection{Releasing Data}
Many InfoSec companies hold their collection of malware samples as proprietary.
As mentioned above, malware authors could also use this information to
thwart existing architectures or solutions built on this data---further risking their clients' systems.
Another consideration is that each collection service may be biased to certain demographics, location, network infrastructure, political ties, etc. that may attract certain types of attacks.
Thus, care should be taken when considering the general applicability of a released dataset.

\subsubsection{Feature Representation}
Distributing live malware samples is a security risk, especially for users who are not accustomed to handling malware.
As a result, most recent datasets first extract predetermined features from a set of malware examples and distribute a dataset of the extracted features.
Many ML papers on detecting malware claim that ML models can go beyond just using signatures and recognizing behaviors.
However, if the features are not capable of identifying behaviors, it is unrealistic to expect ML models to detect behavior.
As we survey below, many features are syntactic and lose the behavioral meaning.

\subsubsection{Obtaining Labels}
Obtaining labels for executables is challenging because malware samples do not broadcast they are malware.
Many of the current datasts use tools like VirusTotal \cite{VT}, which provide the output from multiple antivirus tools, to create labels.
As there is gray area where the antivirus tools disagree, often only samples that are identified as malware by a majority of the tools are labeled as malware and others are discarded.
This provides a biased sample that uses the most obvious examples and is no longer representative of the data that will encountered in real world deployments.
Using the most obvious examples creates easily separable training data and overly optimistic performance expectations as the data does not represent the real-world scenario.
In conjunction with the feature representation, if the labels are poor so will be the resulting ML model.

\subsection{Review of Datasets}

\begin{table*}[ht]
	\begin{threeparttable}
		\caption{Summary of malware datasets used for ML}
		\label{tab:datasets}
		\begin{tabular}{llrlrrrr}
			\toprule
			Dataset & Year & Cite & Representations & \# Samples & Labels & Labeling & Max Acc\\
			\midrule
			\multicolumn{8}{c}{Highly Cited}\\
			\midrule
			VX Heaven \cite{VXHeaven} & 2010 & ? & Live executables & Varies & Varies & Curated & N/A\tnote{1}\\
			VirusShare \cite{virusShare} & 2011 & > 300 & Live executables & Varies & Varies & Curated & N/A\tnote{1} \\
			MalImg \cite{Nataraj2011_MalImg} & 2011 & 417 & Gray-scale images & 9,458 & 25 families & MSSE & 99.80\% \\
			MS Malware Classification \cite{Ronen2018_MSKaggle} & 2015 & 76 & Disassembly and hexadecimal & 10,868 & 9 Families & MSSE & 99.97\% \\
			EMBER \cite{Anderson2018_EMBER} & 2017 & 46 & Parsed and histogram counts & 1,100,000 & Good, Bad, ? & VirusTotal & 99.90\%\\
			MalRec \cite{Severi2018_Malrec} & 2018 & 11 & System calls, memory contents\tnote{2} & 66,301 & 1,270 families & VirusTotal\tnote{3} & N/A\tnote{1}\\
			\midrule
			\multicolumn{8}{c}{Less Cited}\\
			\midrule
			Malware Training Sets \cite{Ramilli2016_MW_dataset} & 2016 & 2 & Counts from analysis reports & 4764 & 4 families & Curated & -\\
			Mal-API-2019 \cite{Catak2019_APIdataset} & 2019 & 1 & System call traces & 7,107 & 8 families & VirusTotal & -\\
			Meraz'18 Kaggle \cite{Kaggle2018} & 2018 & \textasciitilde 1 & Parsed features & 88,347 & Good v Bad & Curated & 91.40\%\tnote{4}\\
			\bottomrule
		\end{tabular}
		\begin{tablenotes}
			\item[1] {\footnotesize There is no established dataset making comparisons between studies difficult.}
			\item[2] {\footnotesize Also provides full system replays of malware execution, however the authors note non-trivial efforts to get them to work on other systems.}
			\item[3] {\footnotesize Uses AVClass \cite{Sebastian2016_RAID} which leverages VirusTotal.}
			\item[4] {\footnotesize Report accuracy on the Kaggle challenge leader board.}
		\end{tablenotes}
	\end{threeparttable}
\end{table*}

There are currently several repositories for ML-based malware detection that either identify malware families or discriminate malware from goodware.
Table \ref{tab:datasets} summarizes several available datasets used for ML.

\subsubsection{Live Malware Repositories}

There are several repositories containing live malware, which pose a threat to inadvertently infecting one's own computer or network as well as providing malicious software to adversaries.
On the other hand, the malware samples provide a valuable resource enabling malware analysis and research.
For example, VX (Virus eXchange) heaven \cite{VXHeaven}, with the mantra: "Viruses don't harm, ignorance does!" seeks to provide information about computer viruses including articles, source code, malware samples, and books to help educate whomever is interested.
There are several similar repositories such as theZoo (a.k.a. the malware DB) \cite{theZoo} and Virus Share \cite{virusShare} for free or Virus Total \cite{VT} which is available for a fee and also contains benign samples.

This is the ideal situation where a researcher has access to the raw data.
Despite having access to the entire malware sample, as discussed previously getting the samples into a format suitable for ML is challenging.
Thus, simple features are often extracted such as metadata from the PE header, imported DLLs, and byte counts (more details on extracted features are given in Section \ref{sec:features}).
These simple features resulted in high detection rates (98.8\%) \cite{Vyas2017}.
Thus, from the ML perspective, there is little reward to working with live malware samples.
With live malware repositories, studies are difficult to compare as each select different subsets of malware samples to analyze and there is no common base publication to trace attribution.
However, the amount of malware samples is impressive.
On Virus Share, there over 34 million samples as of this writing.

\subsubsection{MalImg}
\begin{figure}
	\begin{tabular}{cc}
		\includegraphics[width=0.25\textwidth]{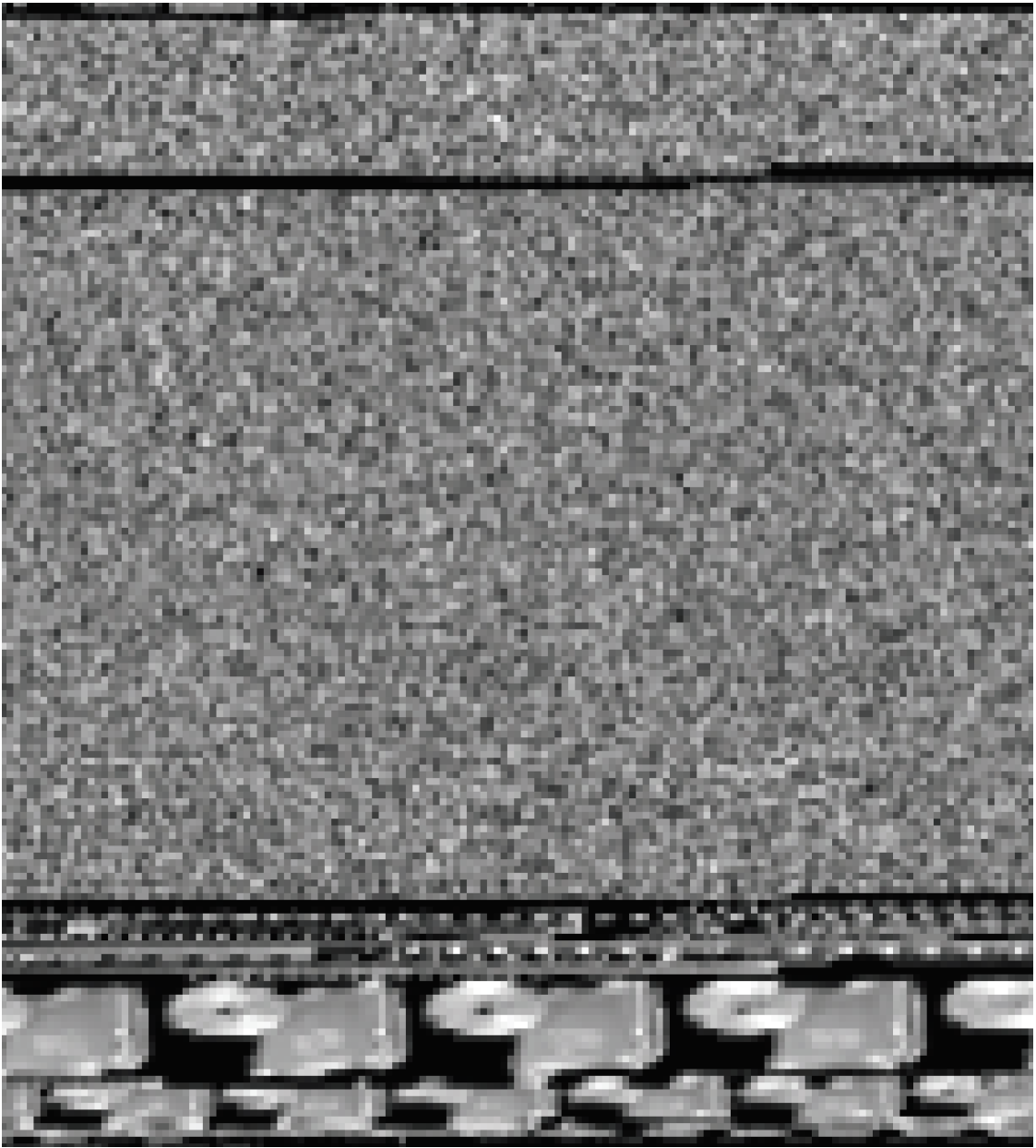} & 
		\includegraphics[width=0.1\textwidth]{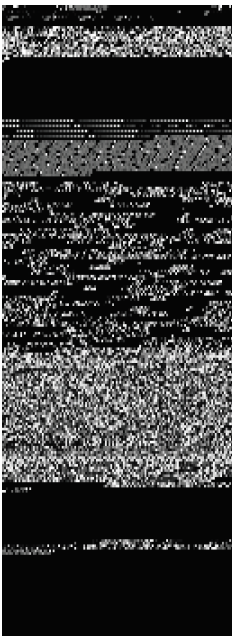}\\
		a & b \\
	\end{tabular}
	\caption{Examples of malware represented as gray-scale images from a) Fakerean and b) Dontovo.A malware families.}
	\label{fig:malimg}
\end{figure}

The MalImg dataset \cite{Nataraj2011_MalImg} was motivated by the success of deep learning (DL) in image processing and represents malware samples as gray-scale images.
The paper is more about a technique, but the dataset used in the study was released and subsequently used in other studies.
In MalImg, binary values from an executable are converted to 8 bit unsigned integers, organized into a 2-dimensional array and visualized as a gray-scale image in the range [0, 255] where 0 is black and 255 is white.
Examples of malware represented as images are shown in Figure \ref{fig:malimg}.
The benefit of this approach is that separate static or dynamic analysis is not required.

The authors observe malware that belongs to the same family are visually similar in layout and texture when visualized as an image.
In their preliminary analysis, the authors extract texture features from the generated gray-scale images using GIST \cite{Torralba2003_ICCV} which extracts frequency and orientation components from an image.
On a dataset with 9,458 malware samples from 25 different families, they get 97.18\% accuracy and 99.2\% when variants of a malware family are combined with a simple 3-nearest neighbor classifier\footnote{A classifier that predicts the majority class of the 3-closest examples}.
Follow up work is better able to discriminate between the family variants achieving accuracy of 98.52\% when using a convolutional neural network \cite{Kalash2018_NTMS} and 99.80\% with principal component analysis and a support vector machine \cite{Ghouti2020_IET}.

From an ML perspective, achieving such high classification accuracy is somewhat concerning as there is fear that the model has either overfit the problem or the problem is easily separable.
Practically, this means that the model will have high generalization error when deployed.
Thus, the dataset might not represent real world conditions well and give unrealistic performance expectations.

\subsubsection{MS Malware Classification}
The Microsoft Malware Classification Challenge \cite{Ronen2018_MSKaggle} was developed as a Kaggle competition to classify malware samples into one of nine malware families.
It was released in 2015 and has since been used in several studies, being cited more than 70 times at the time of this writing.

The hexadecimal representation of the binary content without the PE header as well as meta-information (function calls, op codes, strings, etc.) from the IDA disassembler is provided for each malware sample.
The raw hexadecimal representation allows a user to extract static features that are desired rather than be limited to what is provided.
Current reported performance on the dataset claims 99.70\% \cite{Ghouti2020_IET} and 99.97\% accuracy \cite{Kalash2018_NTMS} using image-based features.

\subsubsection{EMBER}
The Endgame Malware BEnchmark for Research (EMBER) dataset \cite{Anderson2018_EMBER} is a collection of extracted features from 1.1 million executables divided into 900k training and 200k test samples and has emerged as one of the most popular datasets.
EMBER provides features that are consistent with previous work, but what sets EMBER apart from others is the amount of samples.
One of the key benefits of a benchmark dataset is that it enables other researchers who may not be able to gather the data to conduct research in the field.

EMBER has been used in several studies, a few of which we highlight in Tables \ref{tab:EMBER_perf1} and \ref{tab:EMBER_perf2}.
In the original paper, the authors achieved a 98.2\% detection rate with a 1\% false positive rate.
This was further improved to almost perfect detection achieving a 99.4\% detection rate and an AUC value of 0.9997 \cite{Pham2018_FDSE} as shown in Table \ref{tab:EMBER_perf1}.
These results highlight the ``easiness'' of the dataset.
Further highlighting this problem, Vinayakumar et al. \cite{Vinayakumar2019_IEEE} modify a DL  technique aimed at malware detection (MalConv \cite{Raff2017_AAAI}) and achieve nearly perfect accuracy and precision (Table \ref{tab:EMBER_perf2}).
\begin{table}
	\caption{Reported false alarm rate, detection rate, and area under the ROC on the EMBER dataset \cite{Pham2018_FDSE}.}
	\label{tab:EMBER_perf1}
	\begin{tabular}{lccc}
		\toprule
		& False alarm & Detection rate & Area under \\
		Model & rate (FPR) & (TPR) & ROC (AUC)\\
		\midrule
		& 0.10\% & 92.20\% & \multirow{2}{*}{0.9982}\\
		MalConv \cite{Anderson2018_EMBER} & 1.00\% & 97.30\% & \\
		\midrule
		Gradient-based  & 0.10\% & 92.99\% & \multirow{2}{*}{0.9991}\\
		Decision Tree \cite{Anderson2018_EMBER} & 1.00\% & 98.20\% & \\
		\midrule
		& 0.10\% & 97.57\% & \multirow{2}{*}{0.9997}\\
		Pham et al. \cite{Pham2018_FDSE} & 1.00\% & 99.39\% & \\
		\bottomrule
	\end{tabular}
\end{table}

\begin{table}
	\caption{Reported accuracy, precision, recall and F1-score on the EMBER dataset \cite{Vinayakumar2019_IEEE}.}
	\label{tab:EMBER_perf2}
	\begin{tabular}{lcccc}
		\toprule
		Model & Accuracy & Precision & Recall& F1\\
		\midrule
		MalConv \cite{Raff2017_AAAI} & 98.8\% & 99.7 & 97.9 & 98.8\\
		GBDT \cite{Anderson2018_EMBER} & 97.5\% & 99.0 & 96.2 & 97.1\\
		KNN & 95.1\% & 95.5 & 94.6 & 95.1\\
		DT & 96.9\% & 97.1 & 96.7 & 96.9\\
		RF & 97.0\% & 98.6 & 95.3 & 96.9\\
		SVM & 96.1\% & 96.4 & 95.7 & 96.1\\
		DNN & 98.9\% & 99.7 & 98.1 & 98.9\\
		Modified MalConv \cite{Vinayakumar2019_IEEE} & 99.9\% & 99.7 & 100.0 & 99.9\\
		\bottomrule
		
	\end{tabular}
\end{table}

With results this impressive, there is a tendency to suspect either that model has overfit and will not generalize to test data or that the data is too easy.
The authors of EMBER point out that the classes were easy to correctly classify and have attempted to make the task more challenging \cite{Roth2019_harderEMBER} in addition to a number of improvements \cite{Roth2019_CAMLIS}.
The baseline on the updated data is 86.8\%.
Unfortunately, there are few results on the updated dataset.

\subsubsection{Malrec}

Contrary to the other datasets, Malrec provides system-wide traces of malware executions that can be replayed.
It is intended to address the danger of releasing live malware and the limited amount of data that can be collected when running in a sandbox.
The replays capture the state of a system that is executing malware and thus captures the behaviors of malware while not releasing actual malware and provides the ability to retrospectively extract features that were not considered relevant when the malware was first executed.
There are currently 66,301 malware recordings collected over a two year period.
The major downside is the very large size of the data (currently
1.3TB) and the complexity in getting everything set up can be quite challenging versus simply having a dataset that is ready to use.

The authors extracted several datasets from the system-wide recordings including bag-of-word counts for textual data in memory, network activity, system call traces, and counts of data instruction mnemonics.
The authors examined a use case in which they extracted features to use for ML.
They created a word list of all words between 4 and 20 characters long from the English Wikipedia---resulting 4.4 million terms. 
They then monitored memory reads and writes looking for byte sequences that matched words in their list.
Further, terms were removed that appeared in a baseline of running goodware as well as frequent terms that appeared in more than 50\% of the samples and rare terms that appeared in less than 0.1\% of the samples. 
This process resulted in \textasciitilde 460,000 terms.
Finally, Term Frequency Inverse Document Frequency (TF-IDF) scores were used rather than raw counts.
The dimensionality was further reduced to 2048 input features using PCA.
A deep neural network trained on this data achieved a median F1-score of 97.2\% across all of the malware families.

Despite having system-wide information, having a PCA version of the TF-IDF of the text from memory reads were sufficient for their dataset to achieve high accuracy.
While this is positive, it seems to go against common wisdom in InfoSec and presents somewhat of a paradox in the claims of ML and what is observed in deployed systems.
Analyzing the memory contents in a bag-of-words fashion loses context, and we argue, that it is akin to learning a signature.
We conclude that the ML model is able to quickly learn an effective signature-based malware detection system.

\subsubsection{Other Datasets}

As shown in Table \ref{tab:datasets}, there are a handful of other datasets that have been created, often by other security companies and hobbyists \cite{Ramilli2016_MW_dataset, Catak2019_APIdataset, Kaggle2018}.
These datasets have not been widely adopted nor is it apparent how much maintenance they receive.
We include them here for completeness, but they do not provide any feature representations that are not included in the already discussed datasets.
We point out that the Mal-API-2019 \cite{Catak2019_APIdataset} is attractive because system calls, if the context is preserved, should provide more semantic features.
However, system call traces for a large number of malware is provided by MalRec.

\section{Analysis of Datasets and Features}
\label{sec:features}
Given that there are several datasets with different feature representations, we examine which features contribute to the performance of an ML model across the datasets.
We find that 1) the features are overwhelmingly syntactic and very few attempt to extract semantic features \emph{and} are careful to maintain semantic information, and that 2) the most useful features vary across datasets.
These results help confirm that the ML models operate more like signature-based malware detection mechanisms and are not able to pull out salient behavioral information.
The inconsistency in feature importances suggests that the ML model learned some nuances rather than the semantics.
Hence, the ML models should not be expected to detect novel forms of malware based on their behavior as there is a semantic gap between the data and the task.

Raman \cite{Raman2012_features} examined which features are the most discriminative between malware samples from VX Heaven and software that comes installed by default on Windows operating systems.
They were able to achieve a true positive rate of 98.6\% with a false positive rate of 5.7\% by only extracting \emph{seven} features from the files. 
Despite the impressive performance, the selected features are easily manipulable and may be better suited toward discriminating between Microsoft programs and non-Microsoft programs:
\begin{enumerate}
	\item The size of the debug-directory.
	Microsoft-related executables have a debug directory while others may not.
	\item The version of the file.
	This is user defined. Microsoft related executables generally had a larger version while most malware had a image version of zero.
	\item The relative-virtual address of the import address table.
	The value of this feature is 4096 for most clean files and 0 or a very large value for malware.
	\item The size of the export table.
	The size is generally non-zero for clean files and zero for malware.
	\item The size of the resource section.
	Many malware samples have no resources while Microsoft files often do.
	\item The size of the second section in the executable.
	In the dataset, several malware samples only had one section.
	\item The number of sections.
	This relates to the previous feature and the authors are not able to make any clear connections between malware and goodware.
\end{enumerate}
It is clear that the features do not capture the behavior of the malware yet obtain high classification accuracy.
Despite the high performance, the model could easily be defeated by an adversary by including a debug directory or setting the version of the software to be greater than zero.
As the dataset does not represent the real-world problem well, it affects the robustness of an ML model trained on that data.
A high false negative rate would be expected with a larger set of goodware that will be encountered in deployed settings.

Ahmadi et al. \cite{Mansour2016_CODASPY} extracted a large number of features that are commonly used in ML models from the hexadecimal representation and disassembled files from the Microsoft Malware Classification Challenge dataset with the intent of identifying features that are the most discriminative.
The examined features include:
\begin{enumerate}
	\item byte counts (BYTE).
	\item the size of the hexadecimal representation and the address of the first byte sequence (MD1).
	\item byte entropy (ENT).
	\item image representation using Haralick features (IMG1) and Local Binary Patterns (IMG2).
	\item histogram of the length of strings extracted from the hexadecimal file (STR).
	\item the size of, number of line in the disassembled file (MD2).
	\item the frequency of a set of symbols in the disassembled file (-, +, *, ], [, ?, @) (SYM).
	\item the frequency of the occurrence of a subset of 93 of possible operation codes in the disassembled file (OPC).
	\item the frequency of the use of registers (REG).
	\item the frequency of the use of the top 794 Window API calls from a previous analysis of malware (API).
	\item characteristics of the sections in the binary (SEC).
	\item statistics around using db, dw, and dd instructions which are used for setting byte, word, and double word and are used to obfuscate API calls (DP).
	\item the frequency of 95 manually chosen keywords from the disassembled code (MISC)
\end{enumerate}

Table \ref{tab:MS_feat_acc} shows the classification accuracy on the training set and from using 5-fold cross-validation for each subset of extracted features using gradient boosted decision trees.
There are several feature groups that achieve over 99\% accuracy including MISC which simply counts the number of times a hand-selected keyword appears.
None of the features preserve behaviors and most capture syntax.
The system calls get at the behavior, but there is no context.

\begin{table}
	\caption{The accuracy on the training set and using 5-fold cross-validation on the Microsoft Malware Classification Challenge dataset \cite{Mansour2016_CODASPY}.}
	\label{tab:MS_feat_acc}
	\begin{tabular}{cccc}
		\toprule
		 & & Train & 5-CV\\
		Feature & \# Features & Accuracy & Accuracy \\
		\midrule
		\multicolumn{4}{c}{Hexadecimal file} \\
		\midrule
		ENT & 203 & 99.87\% & 98.62\%\\
		BYTE & 256 & 99.48\% & 98.08\%\\
		STR & 116 & 98.77\% & 97.35\%\\
		IMG1 & 52 & 97.18\% & 95.50\%\\
		IMG2 & 108 & 97.36\% & 95.10\%\\
		MD1 & 2 & \textbf{85.47\%} & \textbf{85.25\%}\\
		\midrule
		\multicolumn{4}{c}{Disassembled file} \\
		\midrule
		MISC & 95 & 99.84\% & 99.17\% \\
		OPC & 93 & 99.73\% & 99.07\% \\
		SEC & 25 & 99.48\% & 98.99\% \\
		REG & 26 & 99.32\% & 98.33\% \\
		DP & 24 & 99.05\% & 98.11\% \\
		API & 796 & 99.05\% & 98.43\% \\
		SYM & 8 & 98.15\% & 96.84\% \\
		MD2 & 2 & \textbf{76.55\%} & \textbf{75.62\%} \\
		\bottomrule
	\end{tabular}
\end{table}

Surprisingly, MD1 and MD2 (i.e. file size) achieve about 85\% and 76\% accuracy respectively (random is 11.11\%).
This highlights a concern that there are features which may be discriminative but are easily manipulated.
Thus these features can be used adversarially and do not represent the underlying behavior that is desired to be modeled.

Oyama et al. \cite{Oyama2019_EMBERFeats} examine which features have the largest impact on the EMBER dataset.
EMBER contains several feature groups: 
\begin{enumerate}
	\item General file information from the PE header such as virtual size of the file, thread local storage, resources, as well as the file size and number of symbols.
	\item Header information from the COFF header providing the timestamp, the target machine, linker versions, and major and minor image versions.
	\item Import functions obtained by parsing the address table.
	\item Exported functions.
	\item Section information including the name, size, entropy virtual size and list of strings representing section characteristics.
	\item Byte histogram representing the counts of each byte value.
	\item Byte-entropy histogram approximating the joint distribution of entropy and a given byte value.
	\item String information providing simple statistics about printable strings that are at least five characters long. This feature also specifically provides information on strings that begin with ``C:\textbackslash", ``http://", ``https://" or ``HKEY\_".
\end{enumerate}
Table \ref{tab:EMBER_feats} shows the accuracy for each feature group.
The imports, which also have the largest number of features, has the highest accuracy as 77.8\%.
Oyama et al. also report that header, imports, section, and histogram feature groups (together) provide accuracy up to about 90\% accuracy.
The remaining 2.7\% comes from the other feature groups.
As with the Microsoft Malware Classification Challenge, none of the features include behavioral information.

\begin{table}
	\caption{Reported accuracy and number of features for each feature set in the EMBER dataset \cite{Oyama2019_EMBERFeats}.}
	\label{tab:EMBER_feats}
	\begin{tabular}{lcc}
		\toprule
		Feature set & Number of features & Accuracy\\
		\midrule
		imports & 1280 & 77.8 \\
		section & 255 & 68.2\\
		histogram & 256 & 68.1\\
		byte entropy & 256 & 61.8\\
		strings & 104 & 61.4\\
		general & 10 & 56.0\\
		header & 62 & 52.9\\
		exports & 128 & 17.2\\
		\midrule
		All & 2,351 & 92.7\\
		\bottomrule
	\end{tabular}
\end{table}

Other work makes similar observations on various datasets: 
\begin{itemize}
	\item the PE headers are the most discriminative \cite{Yan2013_DIMVA}
	\item on VX Heaven PE-Miner \cite{Shafiq2009_PE-Miner} achieves a detection rate greater than 99\% only using structural information (PE and section header information), DLLs and object files as features.
\end{itemize}
Further, other work noted that the count features (histograms) used in previous work has promoted overfitting and, in the combination with the labels, produced overly optimistic results \cite{Raff2018_JCVHT}.
This further indicates a change is needed in how the data for ML is represented.

\section{Adversarial ML}
Malware detection is an inherently adversarial domain where attackers constantly try to thwart defenses and defenders try to stay one step ahead of attackers.
We briefly address work in adversarial ML.
Adversarial ML is a broad field which we don't aim to fully cover in this section but highlight existing approaches to evading ML malware detectors.
Some papers have demonstrated moderate success in evading ML classifiers by only changing one pixel, in an image, or one byte, in a malware binary \cite{song2020automatic,su2019one}.
Many papers in the general space assume a neural network architecture which permits specific attack vectors on those models \cite{Szegedy2014_NNprops}.
However, Kucuk and Yan \cite{MalwareAdversarialExamples_2020} demonstrate attacks on a variety of different feature representations that do not rely on a neural network architecture or full knowledge of the classifier.
For any proposed attack, a defense can be constructed but the potential attack surface is vast.
A robust solution to unseen attacks is particularly challenging in this domain and most defenses often come at the expense of accuracy \cite{Wang2017_KDD}.

Evaluating existing feature representations highlights some of the general ambiguity within malware classification.
Prior work attacked two different feature representations (byte code frequencies and system call frequencies) \cite{MalwareAdversarialExamples_2020}. 
Both of these representations, or similar variations, are common in many ML malware classification papers and are used in the reviewed work.
Image representations of malware can be seen as a more sophisticated version of the byte code frequencies.
However, the frequency of individual components doesn't determine the overall behavior of software.
For evading standard signature methods, malware authors insert junk code that isn't anticipated to run or null operators that cause no change of state.
Similarly, insertion of superfluous code yields effective obscuration of malware to many malware machine learning classifiers \cite{park2019generation, demetrio2020efficient}.
Al-Dujaili et al. \cite{Aldujaili2018_adversarial} demonstrate that neural networks can learn the junk code to insert to evade ML-based detectors.
Grosse et al. \cite{Grosse2017_ESORICS} effectively evade detection by an ML-based detector more than 63\% of the time but make the observation that they use static features as perturbing dynamic features is significantly more challenging.
Further, other prior work \cite{Demetrio2019_ExplVulns} showed that when presented as an image, ML did not learn anything meaningful from the text and data sections but learned discriminating features from the header.
Thus, detection could be evaded by simply modifying the syntactic features.
Part of the susceptibility to adversarial attacks arises from not capturing behavioral information.

\section{Behavioral-based Datasets}
\label{sec:behavior}



%
%
%
%

ML models offer the ability to detect novel malware based on their behaviors.
As we have shown, most benchmark datasets do not contain behavioral information or it is lost when extracting features.
As a first step to modeling behaviors, we provide labels for the behavior expressed in malware so the ML model can search for behavioral artifacts.
Extracting behavioral information from an executable is a challenging problem that is a current research area for PA.
We do not claim to solve it here, but offer a process using threat reports corresponding to malware families.

We propose that behaviors consist of a) a high-level description of the intent and b) low-level ``primitives'' that accomplish the respective behavior.
A primitive is a sequence of steps that must occur for the behavior to be successful.
These primitives will vary by representation and malware family/toolkit and may involve multiple systems (e.g., network and host).
Additionally, the sequence will be ordered or partially ordered (i.e., one step depends on the previous step(s)).
It is possible that primitives may contain conditional statements that are represented better by a directed graph than a sequence.
Thus, the feature representation is non-trivial.

Further complicating the issue, multiple primitives may exist that accomplish the same behavior.
For example, persistence is a common behavior for malware as it allows the attacker to maintain a foothold on the machine or network.
To achieve persistence an attacker could copy the malware to the \textit{startup} folder or modify the Registry to execute it whenever the machine reboots.

To label the behaviors, we leverage the MITRE Malware Behavior Catalog
(MBC) \cite{MBC}.
MBC was designed to support malware analysis while mapping into the MITRE ATT\&CK\textsuperscript{\textregistered} knowledge base \cite{attack}.
ATT\&CK documents common tactics, techniques, and procedures that advanced persistent threats use against Windows enterprise networks.
The behaviors are organized according to the \emph{objective} of the malware such as ``Anti-Behavioral Analysis,'' ``Command and Control,'' or ``Persistence.'' 
Each objective contains behaviors and code characteristics (techniques) that support that objective.
For ``Persistence'' some of the techniques include \emph{Application Shimming}, \emph{DLL Search Order Hijacking}, and \emph{Scheduled Task}.
Each technique has an explanation for what it covers and some can belong to multiple objectives---the ``Scheduled Task'' technique could be under the ``Execution,'' ``Persistence,'' or ``Privilege Escalation'' objective.

\begin{table*}[ht]
	\caption{Malware Behavior Label Example for Microsoft Malware Classification Challenge}
	\label{tab:labels}
	\begin{tabular}{r|cc|cccc|cccc}
		\toprule
		{\bf Objective:} &  \multicolumn{2}{|c}{Collection} & \multicolumn{4}{|c}{Credential Access} & \multicolumn{3}{|c}{Defense Evasion} & \dots\\
		\midrule
		\multirow{2}{*}{{\bf Technique:}} & Local & Man in the & & Steal Web & Credential in & Credentials & & Indicator & Process & \\
		 & System & Browser & Hooking & Session & Web Browser & in Files & Masquerading  & Removal & Injection & \dots \\
		\midrule
		{\bf Gatak} & x & - & x & - & - & - & x & - & x & \dots \\
		{\bf Ramnit} & x & x & x & x & x & - & - & - & x & \dots \\
		{\bf Lollipop} & x & - & - & - & - & - & - & - & - & \dots \\
		{\bf Kelihos} & x & - & - & - & - & - & - & - & - & \dots \\
		{\bf Vundo} & x & - & - & - & - & x & x & - & x & \dots \\
		{\bf Simda} & x & - & - & - & - & - & x & - & - & \dots \\
		{\bf Tracur} & - & - & - & - & - & - & - & - & - & \dots \\
		\bottomrule
	\end{tabular}  
\end{table*}


We label the behaviors of a malware family using corresponding open-source threat reports.
The information in the threat reports is then mapped to the ``objectives'' and ``techniques'' outlined by MBC.
In some cases, a judgment has to made about which category is the most appropriate.
To help correct for errors, we label each family multiple times and used a peer-review style to come to conclusions.
The behavioral labels for each family are then extrapolated to individual examples.
The current process is human intensive, subjective, and errors can be made based on variations of a malware family.
Despite these limitations, the behavioral labeling helps align the data to the desired task of identifying novel malware samples through its behaviors.
The labels would allow an ML model to directly learn the behaviors that may be not be discernible using only family information.
Future work could include the use of natural language processing tools to help automate the process.
Additionally, as new malware is analyzed, behaviors could be mapped into the MBC directly bypassing the need for this method.

We label the Microsoft Malware Classification Challenge dataset which includes seven malware families, ($Ramnit$, $Lollipop$, $Kelihos$, $Vundo$, $Simda$, $Tracur$, $Gatak$).\footnote{$Kelihos$ versions 1 and 3 were combined because the threat reports did not distinguish between versions and we dropped Obfuscator.ACY as it was a bucket for obfuscated malware for which the family could not be determined.}
%
%
%
%
%
We document the source and a short definition of the behavior to help develop a repeatable process.
%
%
The result of this process is a hierarchical behavioral labeling of each malware family as shown in Table \ref{tab:labels}.
The compiled version is accessible at \url{https://doi.org/10.6084/m9.figshare.12240980}.
The hierarchical structure captures both the high-level objective and employed technique(s) to meet that objective.
By providing this labeling, an ML model will learn features that are associated with behaviors across all included malware families.
The hope is that by adjusting the objective of the ML algorithms, better features and models can be developed that will improve the deployment of ML-base malware detectors.

We are not the first to suggest the addition of behavioral labels, however our process provides richer behavioral annotation.
Semantic Malware Attribute Relevance Tagging (SMART) \cite{Ducau2019_CAMLIS} uses the output from anti-virus suites and parses keywords from the output to provide additional information.
For example, the output could be \verb|Win32.Virlock.Gen.8| or \verb|TR/Crypt.ZPACK.Gen| and the key words are extracted \verb|Virlock|, and \verb|Crypt| and \verb|ZPACK| respectively.
This provides information that the malware is respectively ransomware and packed.
The keywords align with the objectives in our process but do not provide consistent information on how the behavior is implemented, which our method provides.
The high-level additional information was shown to improve the performance of an ML model \cite{Rudd2019_ALOHA}.
We anticipate similar improved results as well as adjustments in follow-on studies to focus on behaviors.

\section{Directions for Future Research}
Applying machine learning techniques to new areas often requires creatively reformulating problems.
For example, DL techniques are typically better suited to solving statistical problems than performing calculations and working with symbolic data. 
However, by leveraging language models to translate symbolic math problems into suitable formats, DL techniques were used to solve symbolic mathematics problems---outperforming commercial computer algebra systems such as Mathematica or MATLAB \cite{SymbolicMath2019}.
Other non-traditional ML applications seek to generate floor plans from point clouds captured inside of buildings.
By reformulating the problem to predict the center of each room and wall for every point in the cloud, the problem aligns well with existing neural network approaches and allows for interpretable debugging during development \cite{FloorPlan2020}. 

We have shown that current feature extraction and ML techniques optimized for signal processing are inadequate for malware detection.
Future work will require closer collaboration between executable analysis professionals and ML practitioners.
We will need to capture more features such as basic blocks, control flow graphs, etc., that are indicative of behavior and represented such that behavioral information is not lost.
Additionally, improved data will be required that is more representative of the data in deployed scenarios including software that falls in the gray area between obvious malware and goodware.

As new ML and DL methods are developed, some may have more applicability outside of simply classifying malware.
Attention \cite{Bahdanau2014_ICLR,Xu2015_MLR} was introduced as a method to help a DL method focus on the most pertinent portions of the input.
In the InfoSec community, often a piece of software needs to be reverse engineered to some extent to understand the behavior of the software.
Attention allows for an ML model to learn which portions of an executable are the most pertinent to resulting classification \cite{Yakura2019_CompSec}.
Augmented with the behavioral annotations, attention would also indicate which portions of an executable are the most pertinent to that behavior.
This would result in significant decreases in analyst time and potentially lead to improved program understanding.

\section{Conclusions}
In this paper, we reviewed the body of research on providing datasets to train ML models for the classification of malware.
Despite the successful demonstration of classifying malware by ML
models on a number of datasets in academic research, we find that
little ML is used in actual deployments.
We believe that this is due to multiple reasons stemming from a disconnect between the ML and InfoSec communities resulting in a misalignment of questions being asked by the communities and how those are answered.
We have shown that this misalignment in the ML domain stems from a semantic gap in the available data and how that data is represented.
While both communities seek to identify malware, the InfoSec community uses semantic parsing techniques to try to understand what the program is doing and, based on the behavior, a decision can be made to determine if the program is malware or goodware.
The ML community has primarily (inadvertently) examined syntactic features to determine the intent of a program to classify malware.
For ML to make a larger impact in deployed settings, we advocate for 
1) an increased collaboration between the two communities, 
2) improved data in terms of the features that are employed (those that capture behavior), the inclusion of samples that are not clear cut goodware or malware, and the inclusion of behavioral information---modifying the task of determining intent to identifying behaviors, and 
3) the development of a benchmark dataset that more closely aligns with problems encountered by the InfoSec community.
Benchmark datasets have a history of driving and significantly improving ML in a given application area (i.e. computer vision) and an appropriate one could help drive the ML-based malware classification.
As a first step, we proposed a method for annotating datasets with behavioral information and provided behavioral annotations for the Microsoft Microsoft Malware Classification Challenge dataset.

\begin{acks}
This paper describes objective technical results and analysis. Any subjective views or opinions that might be expressed in the paper do not necessarily represent the views of the U.S. Department of Energy or the United States Government.

Sandia National Laboratories is a multimission laboratory managed and operated by National Technology \& Engineering Solutions of Sandia, LLC, a wholly owned subsidiary of Honeywell International Inc., for the U.S. Department of Energy's National Nuclear Security Administration under contract DE-NA0003525. 
SAND2020-4695 C
\end{acks}

\bibliographystyle{ACM-Reference-Format}
\bibliography{lit_review}
\end{document}